\documentclass{mn2e}
\usepackage{psfig}

\def\ltsima{$\; \buildrel < \over \sim \;$}
\def\lsim{\lower.5ex\hbox{\ltsima}}
\def\gtsima{$\; \buildrel > \over \sim \;$}
\def\gsim{\lower.5ex\hbox{\gtsima}}

\def\tobs{\theta_{\rm{obs}}}

\begin{document}

\title[Polarization from Compton drag in GRBs] 
{Compton drag as a mechanism for very high linear polarization in
Gamma-Ray Bursts}

\author[Lazzati, Rossi, Ghisellini, Rees]
{Davide Lazzati$^1$, Elena Rossi$^1$, Gabriele Ghisellini$^2$ \&
Martin J. Rees$^1$ \\ 
$^1$Institute of Astronomy, University of
Cambridge, Madingley Road, Cambridge CB3 0HA, England \\
$^2$Osservatorio Astronomico di Brera. Via E. Bianchi 46, I-23807
Merate, Italy \\ 
{\tt e-mail: lazzati@ast.cam.ac.uk}}

\maketitle

\begin{abstract}
The recent claim by Coburn \& Boggs to have detected a very high
degree of linear polarization in the prompt emission of GRB~021206 has
stimulated interest in how much polarization could arise in gamma-ray
bursts from synchrotron emission. Alternatively, as Shaviv \& Dar have
shown, GRB polarization could be produced by inverse Compton
scattering in the point-source limit. We discuss polarization from a
fireball that upscatters a soft radiation field. We show that, after
the proper angular integration, the residual polarization can be
large, in some cases approaching the point-source limit. We discuss
the probability of realizing the geometrical conditions in which a
large polarization is obtained showing that, for a particularly bright
burst as GRB~021206, the detection of polarization at the first
attempt in the Compton drag scenario is not unlikely.
\end{abstract}


\section{Introduction}

The claimed detection of a very high degree of linear polarization in
the prompt emission of GRB~021206 [$\Pi=(80\pm20)\%$; Coburn \& Boggs
2003] has triggered many theoretical interpretations. Since
polarization is usually associated with some kind of asymmetry in the
way the emitting material is viewed, the possible interpretations can
be divided into two groups. In the first group, discussed in the
observational paper itself, the asymmetry is attributed to a
preferential direction of the magnetic field. Photons are supposed to
be synchrotron emission from a power-law distribution of relativistic
electrons; in the case of a perfectly aligned magnetic field,
polarization can be as high as $\Pi_{\rm{syn}}=(p+1)/(p+7/3)$ where
$p$ is the power-law index of the electron distribution
($n_e(\gamma_e)\propto\gamma_e^{-p}$).  A magnetic field with such an
ordered configuration can not be generated at shocks (Gruzinov \&
Waxman 1999), and should be advected from the engine.  It could be
dominant in the energy budget of the outflow (Lyutikov, Pariev \&
Blandford 2003).

The second possible explanation invokes a particular observer set-up
in order to provide the necessary anisotropy. Waxman (2003) suggested
that a jet with a very small opening angle ($\theta_j\lsim1/\Gamma$)
viewed slightly off-axis ($\theta_o\sim1/\Gamma$ from the jet edge)
would give a polarization level comparable to an ordered magnetic
field even from a shock generated field (see also Gruzinov 1999 for
similar considerations in the afterglow phase). The field should not
be completely tangled, but either parallel to the shock normal (Sari
1999) or contained in the shock plane (Laing 1980; Medvedev \& Loeb
1999), so that relativistic aberration would cause the observer to
look along the edge of the fireball and see therefore an ordered field
(Gruzinov 1999; Ghisellini \& Lazzati 1999).

More detailed analyses of this models (Granot 2003; Nakar, Piran
\& Waxman 2003) showed that the ordered magnetic field can produce a
larger polarization without requiring a particular geometrical
set-up. On the other hand the extreme brightness of the event
($25-100$~keV fluence ${\cal{F}}=1.6\times10^{-4}$~erg~cm$^{-2}$,
Hurley et al. 2002) coupled with the assumption of a fairly typical
redshift\footnote{Unfortunately no optical transient has been detected
for this burst, partly because the burst position was only $18\degr$
from the Sun at the moment of discovery (see Fatkhullin 2003 for a
late upper limit).} $z=1$ yields a narrow opening angle for the jet,
making the probability of random realization of the geometrical set-up
non negligible.

Synchrotron is not the only mechanism able to produce polarized
radiation. In the context of a very narrow hyper-relativistic jet, in
which the point-source approximation is applicable, Shaviv \& Dar
(1995a) discuss polarization as a characteristic signature of inverse
Compton up-scatter (see also a more recent implementation in Dar \& De
Rujula 2003). The same mechanism was considered by Eichler \& Levinson
(2003), who discuss it in the framework of an ensheathed
fireball. They consider the scattering of primary GRB photons,
produced in the inner fast spine of the jet, by electrons advected in
a slower sheath. They obtain a sizable polarization for the observed
radiation under particular observing conditions.

In this paper we consider inverse Compton (hereafter IC) from a
fireball with an opening angle comparable to or larger than the
associated relativistic beaming (Lazzati et al. 2000; see also
Begelman \& Sikora 1987 for a similar scenario in AGNs). In this case
the observed polarization is lower than in the point-source case due
to the fact that the observed radiation comes from different
angles. We numerically compute linear polarization as a function of
the jet opening angle and observer line of sight. This mechanism, with
respect to standard internal shocks, has the advantage of yielding a
large efficiency in the conversion of bulk kinetic energy into
radiation. On the other hand it does not naturally produce a broken
power-law spectrum. A spectrum matching the one observed in GRBs can
be obtained only under appropriate assumptions on the spectrum of the
soft seed photons (Ghisellini et al. 2000).

In \S~2 we describe the geometrical set-up and detail the computation
of polarization from bulk Compton up-scattered photons; in \S~3 we
compute the observed polarization from a fireball as a function of the
geometrical set-up and in \S~4 we discuss our results in the contest
of GRB~021206.

\section{Compton Drag}

Consider an ionized plasma moving relativistically through a photon
field. A fraction $\sim\max(1,\tau_T)$ of the photons ($\tau_T$ is the
Thomson optical depth) suffers inverse Compton scattering on the
relativistic electrons. Their energy is increased in the scattering by
a factor $\sim4\Gamma^2$, where $\Gamma$ is the electron Lorentz
factor. As a net result, due to relativistic aberration, a flow of
high energy photons beamed in an opening angle $\sim1/\Gamma$ is
produced at the expense of the kinetic energy of the plasma flow (see
e.g. Begelman \& Sikora 1987). This radiation mechanism goes under the
name of Compton Drag (hereafter CD) or bulk Compton.

The possibility that the gamma-ray photons in the prompt phase of GRBs
are due to the bulk Compton up-scatter of UV field photons has been
discussed in several geometric setup (Zdziarski, Svensson \& Paczynski
1991; Shemi 1994; Shaviv \& Dar 1995ab; Lazzati et al. 2000; Dar \& De
Rujula 2003). We here concentrate on the case discussed in Lazzati et
al. (2000), in which the relativistic electrons are contained in a
fireball, in contrast to the cannonball (or ballistic) approximation
considered in most of the other works.  Lazzati et al. (2000) showed
that the mechanism can in principle give very high efficiencies,
curing the intrinsic inefficiency of internal shocks (Lazzati,
Ghisellini \& Celotti 1999). In this scenario the shape of the soft
photon field had to be selected in order to reproduce the GRB
spectra. Ghisellini et al. (2000) showed that if the soft photon field
has a power-law distribution of temperatures the resulting spectrum
will resemble that of a GRB. Independently of the details on how the
spectrum is generated, in the electron comoving frame the typical
photon frequency is $h\nu\sim1/1000/\Gamma$~keV~$\ll{}m_ec^2$ and
therefore the scattering is elastic and the polarization properties of
the scattered radiation are independent of frequency. We here show
that, under certain geometrical conditions, it is possible to obtain a
highly polarized flux from CD.

\begin{figure}
\psfig{file=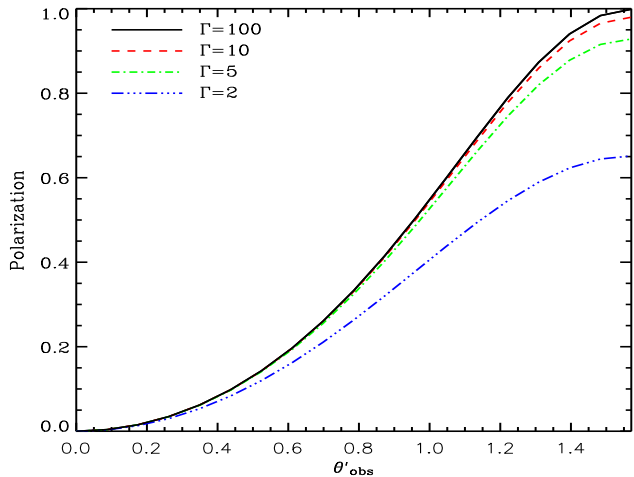,width=\columnwidth,height=6.4cm}
\caption{{Linear polarization as a function of the observing 
angle for a single electron moving relativistically in a uniform
photon field. The observing angle $\theta_{\rm{obs}}$, expressed in
radians, is shown in the electron comoving frame. Electrons with
larger Lorentz factor have more polarizing power since they see, in
their comoving frame, a more anisotropic radiation field. }
\label{fig:sing}}
\end{figure}

Compton (or Thomson) scattered photons are well known to be highly
polarized, with a linear polarization that depends on the photon
scattering angle $\theta$ as (Rybicki \& Lightman 1979):
\begin{equation}
\Pi = {{1-\cos^2\theta}\over{1+\cos^2\theta}}
\label{eq:uno}
\end{equation}

Consider now an electron moving at relativistic speed in an isotropic
photon field. In the electron comoving frame\footnote{Comoving
quantities are primed.} photon aberration makes most of the photons
arrive from one direction. The electron observes a photon field:
\begin{equation}
I^\prime(\nu^\prime)=\delta^3\,I(\nu^\prime/\delta)
\end{equation}
where $\delta\equiv[\Gamma(1-\beta\cos\theta^\prime)]^{-1}$ is the Doppler
factor. The Stokes $Q$ parameter from a single electron can therefore
be computed through:
\begin{equation}
Q=\int_0^{2\pi} \hspace{-2mm}
d\phi^\prime\,\int_0^{\pi} \hspace{-1mm} \sin\theta^\prime\,d\theta^\prime \; \delta^3\,
I\left({{\nu^\prime}\over{\delta}}\right)\,
{{1-\cos^2\theta^\prime_{\rm{sc}}}\over{1+\cos^2\theta^\prime_{\rm{sc}}}}\,
\cos\left(2\phi^\prime_{\rm{sc}}\right)
\label{eq:qsing}
\end{equation}
where the polar and azimuthal scattering angles
$\theta^\prime_{\rm{sc}}$ and $\phi^\prime_{\rm{sc}}$ are specified
through:
\begin{eqnarray}
\cos\theta^\prime_{\rm{sc}}&=&\cos\theta^\prime\cos\theta^\prime_{\rm{obs}}-
\sin\theta^\prime\sin\theta^\prime_{\rm{obs}}\cos\phi^\prime  \nonumber \\
\sin\phi^\prime_{\rm{sc}}&=&{{\sin\theta^\prime\sin\phi^\prime}
\over{\sin\theta^\prime_{\rm{sc}}}}
\end{eqnarray}
and $\theta^\prime_{\rm{obs}}$ is the observing angle with respect to
the electron velocity in the electron comoving frame. Analogously, the
Stokes $U$ parameter can be obtained from Eq~\ref{eq:qsing} by
substituting $\cos\left(2\phi^\prime_{\rm{sc}}\right)$ with
$\sin\left(2\phi^\prime_{\rm{sc}}\right)$. For symmetry reasons, one
always has $U=0$, so that the polarization vector is orthogonal to the
plane containing the electron velocity and the line of sight. The
linear polarization is therefore given by $\Pi=|Q|/I$, where
\begin{equation}
I=2\pi\int_0^{\pi} \hspace{-1mm} \sin\theta^\prime\,d\theta^\prime \; \delta^3\,
I\left({{\nu^\prime}\over{\delta}}\right)
\end{equation}

Eq.~\ref{eq:qsing} can be numerically integrated. The resulting
polarization, as a function of the observing angle
$\theta^\prime_{\rm{obs}}$ is shown in Fig.~\ref{fig:sing} for several
values of the electron Lorentz factor $\Gamma$. In the figure a photon
field $F(\nu)\propto\nu^0$ has been assumed, in order to obtain a
frequency-independent result. The curves in Fig.~\ref{fig:sing} are a
good approximation for any photon field that does not vary sizably
over a factor of $\sim2$ in frequency. The figure shows that the
maximum polarization is obtained for an observer that is at $90\degr$
with respect to the electron velocity vector in the electron comoving
frame. Also, electrons with larger Lorentz factors are more efficient
polarizers, since they see a more anisotropic radiation field due to
relativistic aberration. For $\Gamma\gsim10$ the curves are
indistinguishable and can be approximated with Eq.~\ref{eq:uno} with
an accuracy of $2\%$ for $\Gamma=10$ and $0.1\%$ for $\Gamma\ge100$.

\section{Polarization from a fireball}

\begin{figure}
\psfig{file=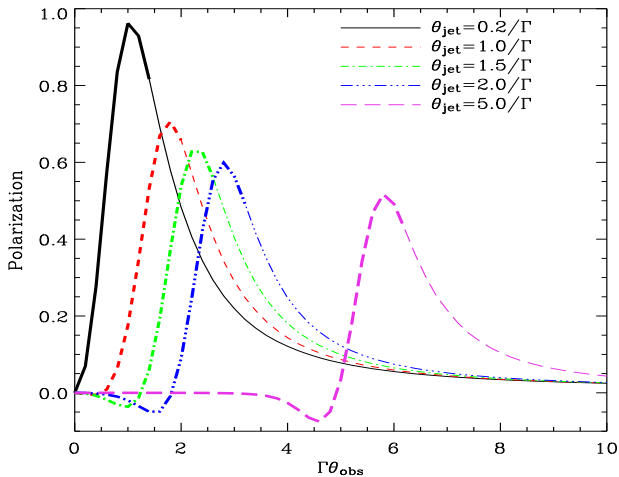,width=\columnwidth,height=6.4cm}
\caption{{Polarization as a function of the observing angle $\theta_o$ 
in units of $1/\Gamma$ for a uniform jet with sharp edges. Different
line styles show the polarization for jets with different opening
angles. The lines are thicker in the region where the efficiency is
larger than $2.5\%$.}
\label{fig:grbdrag}}
\end{figure}

In the previous section we considered the polarization arising from
the interaction of a single electron with an isotropic photon
field. We here consider a jet fireball of opening angle $\theta_j$,
radially expanding at relativistic speed, observed from a viewing
angle $\theta_o$.

In order to compute the polarization that is observed from a fireball,
one has to integrate the single electron polarization over the
fireball surface, taking into account the angle transformations from
the comoving frame to the observed one. Since a comoving observer at
$90\degr$ transforms in an observer that makes an angle $1/\Gamma$ in
the laboratory frame, one expects a maximum in the observed
polarization for an observer located at $1/\Gamma$
from the edge of the jet.

The stokes vector Q (U vanishes due to symmetry considerations as for
the single electron case) is given by:
\begin{equation}
Q=\int_0^{2\pi} \hspace{-2mm} d\phi
\int_0^{\theta_j} \hspace{-1mm} \sin\theta\,d\theta
\,\delta^3\,I^\prime\left({{\nu}\over{\delta}}\right)\,
\Pi(\theta^\prime_{\rm{sc}})\,\cos(2\phi_{\rm{sc}})
\end{equation}
where the Doppler factor $\delta$ must be computed as a function of
the angle between the local velocity and the line of sight. The
scattering angles $\theta_{\rm{sc}}$ and $\phi_{\rm{sc}}$ are as well
defined in the spherical coordinate system where the line of sight is
the polar axis and $\theta_{\rm{sc}}$ is related to the comoving
scattering angle $\theta^\prime_{\rm{sc}}$ through:
\begin{equation}
\cos\theta^\prime_{\rm{sc}} = {{\cos\theta_{\rm{sc}}-\beta}\over
{1-\beta\cos\theta_{\rm{sc}}}}
\end{equation}
Finally, $\Pi(\theta^\prime_{\rm{sc}})$ is the single electron
polarization computed in Eq.~\ref{eq:qsing} (or, for $\Gamma\gsim100$,
the simpler Eq.~\ref{eq:uno}).

\begin{figure}
\psfig{file=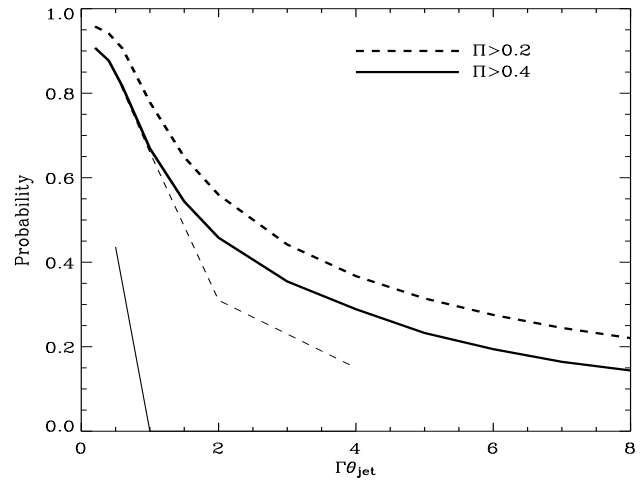,width=\columnwidth,height=6.4cm}
\caption{{Probability for a random observer to detect both a bright 
GRB and a large linear polarization as a function of the jet opening
angle in units of $1/\Gamma$ (thick lines). With thin lines we show
the corresponding probabilities for synchrotron radiation from a
magnetic field predominantly in the shock plane, according to Nakar et
al. (2003). The actual shape of the figure is somewhat dependent on
the threshold levels adopted. Also Compton drag is unable to provide a
large probability for $\Pi>60\%$.}
\label{fig:prob}}
\end{figure}

As for synchrotron models, the observed polarization depends on two
parameters (Granot 2003; Nakar et al. 2003): the opening angle of the
jet $\theta_j$ with respect to its Lorentz factor $\Gamma$ and the
observer off-axis angle $\theta_o$. The resulting polarization, as a
function of the observer angle, is shown for several jet geometries in
Fig.~\ref{fig:grbdrag}, where again the spectral dependence of the
radiation has been neglected for generality. Negative polarization
values correspond to polarization in the plane that contains the jet
axis and the line of sight, while positive polarization values
correspond to an orthogonal polarization angle. The figure shows that
it is indeed possible to obtain a large degree of polarization from CD
in a fireball, especially if the fireball opening angle is close to
the limit $\theta_j\sim1/\Gamma$ (the jet with $\theta_j=0.2/\Gamma$
is shown for completeness). As in the case of synchrotron (Granot
2003; Nakar et al. 2003) the observed polarization is smaller than the
maximum that can be obtained in the non-relativistic case. However,
the net observed polarization from CD is larger, since the maximum
comoving value is $100\%$ rather than $\sim75\%$ for synchrotron.

There are, however, two complications. First, the maximum of
polarization is obtained for observers outside of the jet
($\sim1/\Gamma$ from the jet edge). In this configuration the flux
detected by the observer is smaller than the flux that an on-axis
observer would detect. A highly polarized GRB should therefore appear
dimmer than a non-polarized one, given the same explosion
parameters. We define the efficiency $\epsilon$ as the ratio between
the detected flux and the flux an on-axis observer would detect. Since
GRB~021206 was a rather bright event (Hurley et al. 2002) a situation
in which $\epsilon\ll0.1$ is unlikely. For this reason we have
underlined in Fig.~\ref{fig:grbdrag} with a thicker line-style the
range of off-axis directions in which $\epsilon>0.025$. Even though
the thick lines encompass the peak of polarization, it is clear that,
particularly for the broader jets, only a small fraction of the random
observers would detect a large linear polarization.
In Fig.~\ref{fig:prob}, with thick lines, we show the probability that
a random observer would detect more than a threshold polarization as a
function of the jet opening angle. Threshold polarizations of $20\%$
and $40\%$ have been assumed. For a narrow GRB jet this probability is
not negligible. With a thin line we show, for comparison, the
corresponding probability for synchrotron, according to the
computations of Nakar et al. (2003), for an electron index $p=3$,
corresponding to $\Pi_{\rm{syn}}=75\%$. It is clear that CD can
produce $\sim30\%$ larger linear polarization than synchrotron.

\begin{figure}
\psfig{file=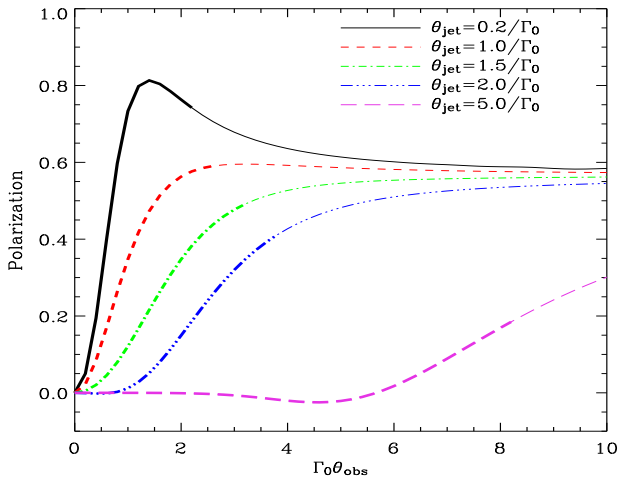,height=6.4cm,width=\columnwidth}
\caption{{Same as Fig.~\ref{fig:grbdrag} but for a Gaussian jet 
(see text). Since in this jet the Lorentz factor is not uniform, the
observing angle $\theta_o$ is shown, in the x axis, in units of
$1/\Gamma_0$, the Lorentz factor along the jet axis.}
\label{fig:gragg}}
\end{figure}

An interesting issue arises if the radiation efficiency of the prompt
GRB phase is large enough to produce a sizable deceleration of the
fireball. This may happen in CD scenarios and even at a larger extent
in the synchrotron case, where only $\sim1/3$ of the dissipated energy
is radiated as MeV photons. The consequences of a deceleration of the
fireball are two: both the jet geometry $\theta_j\Gamma$ and the
observer geometry $\tobs\Gamma$ are affected. In terms of
Fig.~\ref{fig:grbdrag} the location moves to the left on the X-axis,
simultaneously jumping from one curve to another with smaller
$\theta_j\Gamma$. Whether the causes an increase or a decrease of
polarization depends on the starting point, what is certain is that,
if the deceleration is more than a factor of $\sim3\div4$ in $\Gamma$
very small polarization is left.

\subsection{Non uniform jets}

The result we have shown are relative to sharp edged jets, i.e. jets
in which the Lorentz factor and the energy per unit solid angle are
constant inside the cone, dropping to zero (1 for the Lorentz factor)
outside it. A more physical situation should be that of a jet with a
transition region, where the energy per unit solid angle and the
Lorentz factor drop smoothly from their central values to smaller
ones.

\begin{figure}
\psfig{file=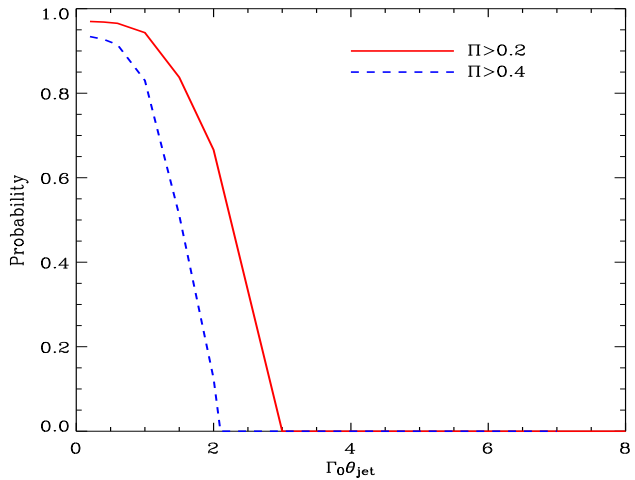,width=\columnwidth,height=6.4cm}
\caption{{Same as Fig.~\ref{fig:prob} but for a Gaussian jet. In this 
case, the probability of observing a large linear polarization is
enhanced in small jets, but quickly drops to zero for
$\Gamma_0\theta_j\gsim3$.}
\label{fig:grob}}
\end{figure}

In Figg.~\ref{fig:gragg} and~\ref{fig:grob} we show the analogues of
Figg.~\ref{fig:grbdrag} and~\ref{fig:prob} in the extreme case of a
Gaussian jet. In this jet, both the Lorentz factor $\Gamma$ and the
energy per unit solid angle ${\cal{E}}\equiv{}dE/d\Omega$ are distributed
according to:
\begin{eqnarray}
\Gamma(\theta) &=& \Gamma_0\,e^{-{{\theta^2}\over{\theta_j^2}}\log2} \nonumber \\
{\cal E} &=& {\cal E}_0 \,e^{-{{\theta^2}\over{\theta_j^2}}\log2}
\end{eqnarray}
so that $\theta_j$ coincides with the half width at half maximum of
the Gaussian. The jet has been truncated at $3\sigma=2.56\theta_j$,
and $\Gamma_0$ has been chosen large enough to ensure that
$\Gamma(\theta)\gg1$. There are two main differences with respect to
the homogeneous jet. First, the polarization outside the core of the
jet remains large out to wide angles, even though the total intensity
drops. This effect is due to the fact that the Lorentz factor in the
edge of the jet is much smaller than that in the core -- which is used
to define the unit on the X-axis of the figure -- and therefore the
observed burst properties vary more slowly with the off-axis
angle. The main difference can be noticed by comparing
Fig.~\ref{fig:grob} with Fig.~\ref{fig:prob}. A narrow Gaussian jet
produces more easily a large linear polarization, but the probability
of observing such a high polarization by a random observer drops to
zero much more quickly with respect to the uniform jet case.

Alternative configurations can be considered. For example, Nakar et
al. (2003) consider a uniform jet with exponential wings only in the
energy per unit solid angle ${\cal{E}}$ and constant Lorentz
factor. What we have shown are the two extreme cases: a pure uniform
jet and a jet with smooth variations of both Lorentz factor and
energy, in order to encompass all possibilities.

\section{Discussion}
\label{sec:disc}

We showed in the previous sections that CD from a moderately narrow
jet (compared to its Lorentz factor $\Gamma$) can produce a large
observable degree of linear polarization. Even though the observed
polarization is larger than that attainable by synchrotron from a
shock generated magnetic field (Granot 2003; Nakar et al. 2003), the
condition $\Gamma\theta_j\lsim5$ must be satisfied in order to have a
non vanishing probability of observing linear polarization at first
attempt. Unfortunately it is not possible to measure either the
Lorentz factor of the fireball or its opening angle, since we do not
have afterglow observations for GRB~021206.  Nevertheless an estimate
of $\theta_j$ can be made thanks to two correlations recently
discovered in the sample of well-observed GRBs. First, we use the
standard beaming corrected energy for GRBs $E\simeq5\times10^{50}$~erg
(Frail et al. 2001). Second, we adopt the ``Amati correlation'', which
relates the isotropic equivalent energy output of the GRB in
$\gamma$-rays with the peak frequency (in $\nu{}F(\nu)$
representation) of its spectrum (Amati et al. 2002). For the case of
GRB~021206 the observed peak frequency is $h\nu_{\rm{peak}}\sim1$~MeV
(Hajdas et al. 2003). Combining this information with the fluence of
the GRB ($25-100$~keV fluence
${\cal{F}}=1.6\times10^{-4}$~erg~cm$^{-2}$, Hurley et al. 2002) we
obtain $z\sim1.5$ and $\theta_j\sim2\degr$, making the jet of
GRB~021206 one of the narrowest detected\footnote{It should be noted
that in deriving the numerical values we have neglected the small
efficiency implied by the measured polarization. Should it be taken
into account, an even smaller opening angle would be derived
(e.g. Nakar et al. 2003).}, yet comparable to the jets of GRB~990123,
GRB~990510 and GRB~000926 (Frail et al. 2001).  The Lorentz factor of
the fireball can be estimated from the observed peak frequency. If the
soft photon field is produced by a simultaneously exploding supernova
(Lazzati et al. 2000), the observed peak frequency is given by:
\begin{equation}
h\nu_{\rm{peak}}\simeq10\,\Gamma^2\,k\,T_{\rm{ph}}
\end{equation}
where $T_{\rm{ph}}\sim10^5$~K is the black body temperature of the
photon field. Inverting the equation we obtain $\Gamma\sim130$. This
yields, as a most probable configuration for GRB~021206:
\begin{equation}
\Gamma\theta_j\sim5
\end{equation}

We conclude that, if the jet configuration of GRB~021206 were
sufficiently sharp-edged, polarization from CD would be a viable
explanation for the RHESSI observation, even though the actual
jet-observer configuration would be somewhat lucky. Under similar
conditions, synchrotron radiation from a shock generated field, would
produce somewhat smaller polarization.

Even though synchrotron radiation from an ordered toroidal magnetic
field component advected from the inner engine is an alternative good
explanation for the detected high level of polarization, we have shown
that the detected polarization does not necessarily imply synchrotron
as the emission mechanism (see also Eichler et al. 2003). CD has also
the advantage of naturally providing a large efficiency, while it
requires some ad-hoc assumptions in order to reproduce the observed
spectrum. There are several ways to discriminate between the two
mechanisms with future observations. First, CD requires a particular
geometrical set-up to produce a large polarization. As a consequence,
only a small fraction of GRBs should be highly polarized, the more
luminous being more likely to be polarized than the dim ones, since
they have narrower jets. On the other hand, if polarization is due to
a toroidal magnetic field advected from the central source, it should
be a general feature of all GRBs, irrespective of their observing
angles and apparent luminosities\footnote{Only the small fraction of
GRBs observed at $\theta_o<1/\Gamma$ from the jet axis should be
unpolarized.}. Secondly, CD polarization is a property of the prompt
event only, and the optical flash should be less polarized, while it
should be highly polarized in the synchrotron case. Finally,
synchrotron polarization has a well defined relation to the spectral
slope of the radiation (Granot 2003; Lazzati et al. 2003 in
preparation), while CD polarization should be almost independent on
the spectral properties of the observed radiation.

\section*{Acknowledgements}
We thank Annalisa Celotti and David Eichler for useful discussions. DL
acknowledges support from the PPARC postdoctoral fellowship
PPA/P/S/2001/00268. ER thanks the Isaac Newton Fellowship and PPARC
for financial support.


\begin{thebibliography}{}
\bibitem{} Amati L. et al., 2002, A\&A, 390, 81
\bibitem{} Begelman M. C., Sikora M., 1987, ApJ, 322, 650
\bibitem{} Coburn W. \& Boggs S. E., 2003, Nature, 423, 415
\bibitem{} Dar A. \& De Rujula A., 2003, MNRAS submitted 
	(astro-ph/0308248)
\bibitem{} Eichler D. \& Levinson A., 2003, ApJ in press
	(astro-ph/0306360)
\bibitem{} Fatkhullin T, 2003, GCN \#2341
\bibitem{} Frail D. A. et al., 2001, ApJ, 562, L55
\bibitem{} Ghisellini G. \& Lazzati D., 1999, MNRAS, 309, L7
\bibitem{} Ghisellini G, Lazzati D., Celotti A. \& Rees M. J., 2000,
	MNRAS, 316, L45
\bibitem{} Granot J., 2003, ApJ, 594, L83
\bibitem{} Gruzinov A., 1999, ApJ, 525, L29
\bibitem{} Hajdas W., Wigger C., Aerzner K.,  Eggel Ch., Guedel M.,
	Zehnder A., Smoth D., 2003, poster presented at the conference
	``Particle acceleration in astrophysical objects'', June 24-28
	2003 ({\tt http://www.oa.uj.edu.pl/konferencje/proc0.html})
\bibitem{} Hurley K. et al., 2003, GCN \#1727
\bibitem{} Laing R. A., 1980, MNRAS, 193, 439
\bibitem{} Lazzati D., Ghisellini G., Celotti A., 1999, MNRAS, 309, L13
\bibitem{} Lazzati D., Ghisellini G., Celotti A. \& Rees M. J., 2000,
	ApJ, 529, L17
\bibitem{} Lyutikov M, Pariev V. I., Blandford R. D., 2003, ApJ
	submitted (astro-ph/0305410)
\bibitem{} Medvedev M. V. \& Loeb A., 1999, ApJ, 526, 697
\bibitem{} Nakar E., Piran T. \& Waxman E., 2003, JCAP submitted 
	(astro-ph/0307290)
\bibitem{} Rybicki G. B. \& Lightman A. P., 1979 {\it Radiative processes in
  	astrophysics}, Wiley Interscience, New York
\bibitem{} Sari R., 1999, ApJ, 452, L43
\bibitem{} Shaviv N. J. \& Dar A., 1995a, ApJ, 447, 863
\bibitem{} Shaviv N. J. \& Dar A., 1995b, MNRAS, 277, 287
\bibitem{} Shemi A., 1994, MNRAS, 269, 1112
\bibitem{} Waxman E., 2003, Nature, 423, 388
\bibitem{} Zdziarski A. A., Svensson R., Paczynski B., 1991, ApJ, 
	366, 343
\end{thebibliography}
\end{document}